\documentclass[preprintnumbers,amsmath,amssymb]{revtex4}
\usepackage{epsfig}
\usepackage{graphicx}
\usepackage{dcolumn}
\usepackage{bm}

\begin{document}

\hfill{CAB-lcasat/04031}
\title{Numerical integration of the discrete-ordinate radiative transfer equation in strongly non homogeneous media.}

\author{M.-P. Zorzano }
\email{zorzanomm@inta.es}
\homepage{http://www.cab.inta.es}
\affiliation{Centro de Astrobiolog\'{\i}a (CSIC-INTA),
        Carretera de Ajalvir km 4,    28850 Torrej\'{o}n de Ardoz, Madrid, Spain}

\author{A. M. Mancho}
 \email{A.M.Mancho@imaff.cfmac.csic.es}
\affiliation{Instituto de Matem\'aticas y F\'{\i}sica Fundamental,
Consejo Superior de Investigaciones Cient\'{\i}ficas, Serrano 121, 28006 Madrid, Spain}

\author{L. V\'azquez}
\email{lvazquez@fdi.ucm.es}

\affiliation{Departamento de Matem\'{a}tica Aplicada. Facultad de
    Inform\'{a}tica. Universidad Complutense. 28040 Madrid}
\affiliation{Centro de Astrobiolog\'{\i}a (CSIC-INTA),
        Carretera de Ajalvir km 4,    28850 Torrej\'{o}n de Ardoz, Madrid,
    Spain}
\date{\today}

\begin{abstract}

We consider the radiation transfer problem in the
discrete-ordinate, plane-parallel approach. We introduce two benchmark problems with exact known solutions and show that for strongly non-homogeneous media the homogeneous layers approximation
can lead to errors of $10\%$ in the estimation of the
intensity.  We propose and validate a general purpose numerical method that
transforming the two-boundary problem into an initial boundary problem, using an  adaptative step integration and an interpolation of the local
optical  properties,  can improve the  accuracy of the solution up to two orders of
magnitude. This is furthermore of interest for
practical applications, such as atmospheric radiation transfer, where the scattering and absorbing properties
of the media vary strongly with height and are only known, based on measurements or models, at
certain discrete points.

\end{abstract}
\maketitle
\section{Introduction}

The mathematical modeling of radiative transfer in which the phenomena of
absorption, emission and scattering are taken into account is usually made
using the linearized Boltzmann equation, also known as radiation transfer
equation. This equation describes the transfer of radiation, with
a given wavelength, through a medium with certain absorbing and
scattering properties.  A particular application of this equation is the study
of radiation transfer in the atmosphere, where the medium properties vary
strongly with height. Getting accurate solutions of this problem is important for evaluating energy balance on planetary
atmospheres as well as for the
so called  "inverse problem" where the boundary conditions (in particular the
ground albedo properties) are deduced from measurements of radiation and
knowledge of the medium characteristics.

In this work we consider the time-independent, monochromatic radiative
transfer equation using the well-tested and widely used discrete-ordinate
method of Stamnes {\sl et al.} \cite{Stamnes} and the plane parallel
approach where the optical properties depend only on the vertical coordinate
$z$. The procedure requires the solution of a system of $n$ 
 coupled linear ordinary differential equations (one for each stream or
 discrete-ordinate component of the intensity). This set of equations is subject to a
 two-point boundary condition at the top and bottom of the medium. In the
 general case no analytic solutions exist for this problem since the medium optical properties (phase function, absorption and
scattering coefficients) depend on the position $z$ in the vertically inhomogeneous medium.

To obtain a formal solution, the medium is generally assumed to be layered with
piecewise constant optical properties, i.e. it is
divided into $N$ adjacent homogeneous layers where the eigenvalues are computed. The coefficients of the solution are determined
by imposing  the continuity condition at the boundaries between adjacent
layers and the two-point boundary conditions at the top and bottom of the
medium \cite{Liou,Kylling}.

Unfortunately in many cases the optical properties of the
 medium are not homogeneous, in fact they show strong variations along the
 vertical axis and have therefore different characteristic length scales. In these cases, as we will see below, the homogeneous layers
 assumption may lead to errors of $10\%$ in the estimation of the scattered radiation. Furthermore in most practical applications the medium characteristics are only known,  based on measurements or calculations, at a discrete number of points \cite{Carmen&Ana}. It is therefore
 required to implement a method that is able to cope both with the strong
 variations on the medium properties and with the discrete character of the
 information available.

Here we solve the discrete-ordinate approximation to the radiative
 transfer equation with a different approach. Since the equations and the boundary conditions are linear in the intensity, the two-point boundary
 problem can be solved with a shooting method.  The problem is then reduced to the
 solution of an initial value problem which can be solved numerically.
 We propose a numerical integration of the initial value problem that applies an adaptive step method
 (such as {\sl step doubling}), interpolates the optical
 properties from the discrete set of available data (for example using a cubic
 spline) and uses a weighted  evaluation of the integrand along the interval (in our case
a 5th order Runge-Kutta scheme). This allows the adaptation of the numerical integration to the
local characteristic length-scales of the problem under consideration. The
 assumption of homogeneity is thus not required. 

As an example we apply this method to solve the
 two-stream  discrete ordinate version of the, non-emitting, radiative
 transfer equation in one spatial dimension (these results can be equivalently extended to multi-stream and emitting
 versions of this equation). We present two benchmark cases with known analytical solutions
 and compare the performance obtained when the step doubling interpolating
 method is used and when the piecewise homogeneity is imposed.  We will show that using the same available discrete information, our method can significantly improve the accuracy of the solution.
In section \ref{RTE} we will first introduce the radiative transfer equation
and the equations for the direct and diffuse intensity components in the
two-stream discrete ordinate approach.
In section \ref{examples} we will describe two benchmark cases characterized
 by linear and  exponential height dependence of the optical coefficients in
the equations. In section \ref{comparison} we will explain our numerical
procedure and will validate our results and those obtained under the
assumption of piecewise homogeneous layers against the exact analytical
 solutions of the benchmark problems.
Finally conclusions are presented  in section \ref{conclusion}.

\section{Multiple Scattering and the Radiative Transfer Equation}\label{RTE}

\begin{figure}
\begin{center}
\includegraphics[width=0.75 \textwidth]{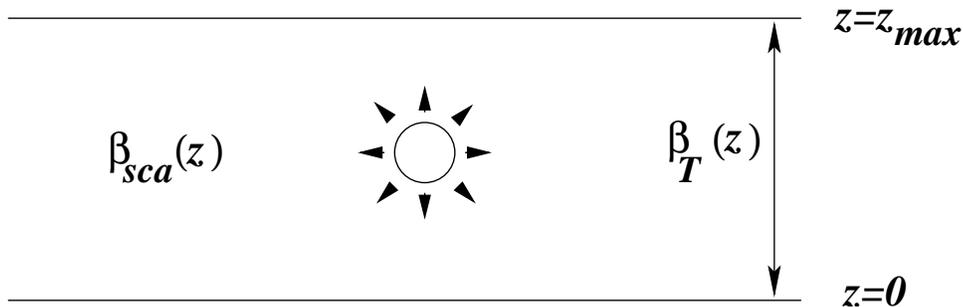}
 \end{center}

\caption{\label{layer}Layer with altitude dependant extinction and scattering coefficients.}
\end{figure}

Our aim is to solve  the radiative transfer problem in the plane parallel approach. As represented in Figure \ref{layer} this corresponds to the case in which optical properties only depend on altitude $z$ in a plane parallel geometry. Following for instance \cite{Stamnes} the radiative transfer  equation in this approach is written as,
\begin{eqnarray}\label{radtr} 
\cos{\theta}\frac{d
  I(z,\theta,\phi,\lambda)}{dz}=-\beta_T(z,\lambda)I(z,\theta,\phi,\lambda)+\frac{\beta_{sca}(z,\lambda)}{4\pi}\int_0^{2\pi}d\phi
  ' \\ \nonumber
\int_0^{\pi} \sin{\theta '}d \theta ' p(z,\theta,\phi,z',\theta  ',\phi ',\lambda)I(z,\theta  ',\phi ')
\end{eqnarray} 
where $\theta$  and  $\phi$ are respectively the polar and azimuthal angles.
Here $\beta_{sca}$ represents the attenuation due to scattering effects and $\beta_T(z,\lambda)$
is the extinction coefficient defined as
$\beta_T(z,\lambda)=\beta_{sca}(z,\lambda)+\sum_i\beta_{i}(z,\lambda)$ where
the summation in $i$ extents to all the molecular components considered. Here
 $\beta_{i}(z,\lambda)=n_i(z)\sigma_i(\lambda)$ is the absorption coefficient of a given component,
$\sigma_i(\lambda)$ is the attenuation cross section due to absorption and
$n_i(z)$ the atmospheric number density for species $i$ (which generally
depends exponentially on the height $z$).
In Eq. (\ref{radtr}), $p(z,\theta,\phi,z',\theta  ',\phi ',\lambda)$ is the
phase function of the scattering particles which is normalized as follows:
$\frac{1}{4\pi}\int_0^{2\pi}d\phi\int_{0}^{\pi}p(\theta,\phi,z',\theta
',\phi',\lambda) \sin(\theta) d\theta=1$. This function gives the probability for a photon of wavelength
$\lambda$, incident on the scattering particle with angles $(\theta  ',\phi ')$ to be
scattered in the direction $(\theta  ,\phi )$ and satisfies 
$p(\theta,\phi,\theta  ',\phi ',\lambda)=p(\cos{\Theta},\lambda)$
where $\Theta$ is the scattering angle, related to the polar and azimuthal
angles by
$\cos{\Theta}=\cos{\theta  '}\cos{\theta}+\sin{\theta '}\sin{\theta}\cos(\phi-\phi ')$.

The solution of Eq. (\ref{radtr}) is splitted into two terms

\begin{eqnarray}
I(z,\theta,\phi,\lambda)=I^{dir}(z,\lambda)+I^{dif}(z,\mu,\phi,\lambda).
\end{eqnarray}
$I^{dir}(z,\lambda)$,  the direct intensity, is the solution of Eq. (\ref{radtr}) when
there is no multiple scattering (no integral term):
\begin{equation}\label{radtrdir}
\mu_0 \frac{d I^{dir}(z)}{d z}=-\beta_T(z) I^{dir}(z)
\end{equation}
where $\mu_0$ is the cosine of the polar angle for the incident radiation.
The diffuse intensity is the solution of

\begin{eqnarray}\label{radtrdif}
\mu\frac{d  I^{dif}(z,\mu,\phi,\lambda)}{dz}=-\beta_T(z,\lambda)I^{dif}(z,\mu,\phi,\lambda)+\frac{\beta_{sca}(z,\lambda)}{4\pi}\int_0^{2\pi}d\phi ' \\ \nonumber
\int_0^{\pi} d \mu ' p(z,\mu,\phi,z',\mu  ',\phi
',\lambda)I^{dif}(z,\mu  ',\phi ')\\ \nonumber
-\frac{\beta_{sca}(z,\lambda)I^{dir}(z,\lambda)}{4\pi}p(z,\mu,\phi,z',-|\mu
_0| ,\phi_0,\lambda)',\lambda).
\end{eqnarray}
with $\mu=\cos{\theta}$. In this equation the variable $\mu$ takes values in the range $-1<\mu <1$. Negative $\mu$ corresponds
to radiation going downwards, whereas positive $\mu$ describes radiation going upwards.
In order to solve Eq. (\ref{radtrdif}) the diffuse intensity is expanded in a
$2n$ Fourier cosine series (from now on we drop the superscript $dif$): $I(z,\mu,\phi,\lambda)=\sum_{m=0}^{2n-1}I^{m}(z,\mu)\cos{m(\phi_0-\phi)}$.
The phase function is expanded in a basis of $2n$ Legendre polynomials $p(z,\mu,\phi,z',\mu ',\phi ',\lambda)=p(z,\cos{\Theta},\lambda)=\sum_{l=0}^{2n-1}(2l+1)g_l P_l(\cos{\Theta})$.
With these transformations and the theorem of addition of spherical harmonics
Eq. (\ref{radtrdif}) becomes a set of integro-differential equations depending
only on the $z$ and $\mu$ coordinates:
\begin{eqnarray}\label{eq0}
\mu \frac{d I^{m}(z,\mu)}{d z}&=&-\beta_T(z,\lambda)I^{m}(z,\mu)+J^{m}(z,\mu),
\end{eqnarray}
where
\begin{eqnarray}\label{J}
J^{m}(z,\mu)&=&\frac{\beta_{sca}(z,\lambda)}{2}\sum_{l=0}^{2n-1}(2l+1)g_l^{m}P_l^{m}(\mu)\\
&\ & \cdot  \nonumber 
\left(
  \int_1^{1} P_l^{m}(\mu ')I^{m}(z,\mu')d\mu  '+\frac{I^{dir}(z,\lambda)}{2\pi}(2-\delta _{0,m})(-1)^{(l+m)}P_l^{m}(|\mu _0|)
  \right)
\end{eqnarray}
being $P_l^{m}(\mu)$  the associated Legendre polynomial, 
$g_l^m=g_l\frac{(l-m)!}{(l+m)!}$, and  $g_l=\frac{1}{2}\int_1^1p(\cos{\Theta})P_l(\cos{\Theta}) d(\cos{\Theta})$.

 In the discrete ordinate
approximation the angular integral term in  Eq. (\ref{J}) is represented by a summation over $n$ Gaussian quadrature points $\mu_s$
(fixed angles) also called "streams". 
The intensity given by  Eq. (\ref{eq0}) must satisfy boundary conditions at the
top ($z=z_{max}$) and bottom ($z=0$) of the medium. Therefore we end up with a system
of $n$ coupled ordinary differential equations of the type (\ref{eq0}), one for each
stream $\mu_s$, and subject to a two-point boundary condition.  
Solving these
equations, we will get a discrete approximation to the angular distribution of
$I^{m}(z,\mu)$ from the top of the
atmosphere to the surface level.

In this work, for simplicity, we will
consider the two-stream approximation to this problem and obtain the two
streams $I^{m}(z,\mu_1)$ (downwards) and $I^{m}(z,\mu_2)$ (upwards) fulfilling the
 boundary conditions that impose, first,  no diffuse
radiation incident at the top, and second, no radiation reflected
back at the surface. This is expressed as follows,

\begin{eqnarray}\label{bc1}
I^{m}(z=z_{max},\mu_1)&=&0\\\label{bc2}
I^{m}(z=0,\mu_2)&=&0.
\end{eqnarray}

In some cases it is preferred to use the optical depth of the medium $\tau$
instead of the vertical geometric distance $z$. This is a non-dimensional
variable which is defined as $\tau(z)=\int_z^\infty \beta_T(z)dz$. It
describes the attenuation within the
medium of an incident beam of radiation  when emission and multiple scattering are ignored, i.e. it is a
measurement of the direct component attenuation.

\section{The benchmark problems}\label{examples}

In this section we define three benchmark problems with an exact solution where
the optical properties $\beta_T(z,\lambda)$ and $\beta_{sca}(z,\lambda)$ in
Eq. (\ref{radtr}) have different altitude ($z$) dependences. 
For the sake of simplicity we consider that particles scatter radiation
uniformly, {\em i.e.} the scattering phase function is $p(\cos \Theta)=1$ and
that $z_{max}=1$. 

For the integral part in  Eq. (\ref{J}) and following the two-stream
approximation we consider a weighted summation over the streams
$\mu_1=-1/\sqrt(3)$ (down) and $\mu_2=1/\sqrt(3)$ (up) which are the zeroes of the second order Legendre polynomial, {\em
  i.e.} $P_2(\mu_i)=0$, $i=1,2$. Then, Eq. (\ref{eq0}) can be transformed in two coupled ordinary  differential equations 
for the down ($I_1$) and up ($I_2$)  radiation intensity:
\begin{eqnarray}
\mu_1 \frac{d I_1}{d z}&=&-\beta_T(z) I_1 + \frac{0.5 \beta_{sca}(z)}{2}\left( I_1+ I_2+ \frac{I^{dir}(z)}{2 \pi} \right),\\
\mu_2 \frac{d I_2}{d z}&=&-\beta_T(z) I_2 + \frac{0.5 \beta_{sca}(z)}{2}\left( I_1+ I_2+ \frac{I^{dir}(z)}{2 \pi} \right).
\end{eqnarray}
 After a change of variables $z'=1-z$ (from now on dropping the prime in $z$) and
taking into account that $\mu_1=-\mu_2$ the above equations may be rewritten as, 

\begin{equation}
\mu_0 \frac{d I^{dir}}{d z}=\left[A(z)-B(z)\right] \mu_2 I^{dir}
\end{equation}

\begin{eqnarray}\label{diff1}
\frac{d I_1}{d z}&=&A(z) I_1 + B(z)  I_2+ I^{dir}(z) \frac{B(z)}{2 \pi} \\\label{diff2}
\frac{d I_2}{d z}&=&-A(z) I_2 - B(z) I_1- I^{dir}(z) \frac{B(z)}{2 \pi} 
\end{eqnarray}
where

\begin{eqnarray}
A(z)&=&\frac{-\beta_T (z)+ 0.25 \beta_{sca}(z)}{\mu_2}\label{eqaz}\\
B(z)&=&\frac{0.25 \beta_{sca}(z)}{\mu_2} \label{eqbz}\\
I^{dir}(z)&=&I_0 e^{\int -\frac{A(z)-B(z)}{\mu_0} \mu_2 dz}
\end{eqnarray}
and $I_0$ is fixed by the condition at the top $I^{dir}(z=0)$. In the following examples
we will solve this problem with the two-point boundary conditions given in Eqs. (\ref{bc1})-(\ref{bc2}),
which  are now explicitely written as $I_1(z=0)=0$ (at the top) and
$I_2(z=1)=0$ (at the bottom). The 
incident intensity on the top is characterized by $I_0=100$ and
$\mu_0=-0.788$. We will consider two different height dependences of the
optical properties.

\subsection{Linear dependence}

We propose as  study-case a linear dependence on the extinction
coefficients. In particular we have considered the functions $A(z)$ and $B(z)$,
\begin{eqnarray}
A(z) &=& {\displaystyle \frac {z\,(4\,a^{2}+c^{2})}{2\,
c}} \\
B(z)&=&{\displaystyle \frac {z \left( 4\,{a}^{2}-{c}^{2} \right) }{2c}}.
\end{eqnarray}
With this choice the general solutions of Eqs. (\ref{diff1})-(\ref{diff2}) for the diffuse components are,

\begin{equation}
{I_1}(z) = e^{(-z^2\,a)}\,{C_1} + \,e^{(z^2\,a)}\,{C_2}  -
 {\displaystyle \frac {I_0\,(\mu_0^{2}-\mu_0 \mu_2)\,(4\,a^{2} - c^{2})}{16\,\pi \,( - c^{2}\,{\mu_2}^{2} + 4\,a^{2}\,{\mu_0}^{2})}} \,e^{\left(-\frac {c\,z^{2}\,{\mu_2}}{2\,{\mu_0}}\right)}\end{equation}
for the downwards radiation, and

\begin{equation}
I_2(z) =  - {\displaystyle \frac {2\,a +c}{2\,a - c}} \,e^{( - z^{2}\,a)}\,
\mathit{C_1} -   {\displaystyle \frac {2\,a-c}{2\,a + c}}  \,e^{(z^{2}\,a)}\,\mathit{
C_2} \mbox{}  - {\displaystyle \frac {I_0\,(\mathit{\mu_0}^{2}+ \mathit{\mu_0
}\,\mathit{\mu_2})\,(4\,a
^{2} - c^{2})}{16\,\pi \,( - c^{2}\,\mathit{\mu_2}^{2} + 4\,a^{2}
\,\mathit{\mu_0}^{2})}}   \,e^{\left(-
\frac {c\,z^{2}\,\mathit{\mu_2}}{2\,\mathit{\mu_0}}\right)} 
\end{equation}
for the upwards component. Here $C_1$ and $C_2$ are arbitrary constants fixed with the boundary
conditions in Eqs. (\ref{bc1})-(\ref{bc2}). The direct component of the intensity is now
\begin{equation}
I^{dir}(z)=I_0\,e^{\left(-\frac {c\,z^{2} \mu_2}{2\,\mathit{\mu_0}}\right)}.
\end{equation}

For the specific choice $c=-10.5, a=4$  the
constants $C_1$ and $C_2$ that satisfy the boundary conditions are $C_1
=-33.0927$ and $C_2 =-6.71346e-05$. For this
particular set of parameters the total optical path is $\tau\approx 3$. The optical functions
$\beta_T(z)$ and $\beta_{sca}(z)$ are shown in Fig. \ref{beta}-I and in Fig. \ref{beta}-II the
attenuation of the direct
intensity as a function of height. The two-stream components of the
diffused intensity are shown in Fig. \ref{exact}-I.

\subsection{Exponential dependence}

In atmospheric layers the gas density typically depends on the altitude as
a growing exponential  from $z=0$ (the top) to $z=1$ (the bottom). For this
reason the choice of exponential dependence in the extinction coefficients is
very convenient as it allows us to validate our method with solutions similar to
those appearing in atmospheres, where our method has been applied \cite{Carmen&Ana}. In particular we have chosen,

\begin{eqnarray}
{A}(z) &=& \,{\displaystyle 
\frac {e^{(a\,z)}\,(a^{2}\,b^{2}+c^{2})}{2c}} \\
{B}(z) &=& \,{\displaystyle \frac {e^{(a\,z)}\,
(a^{2}\,b^{2} - c^{2} )}{2c}} 
\end{eqnarray}

\begin{figure}
\begin{tabular}{cc}

    (I) & (II)  \\
\includegraphics[width=0.45 \textwidth]{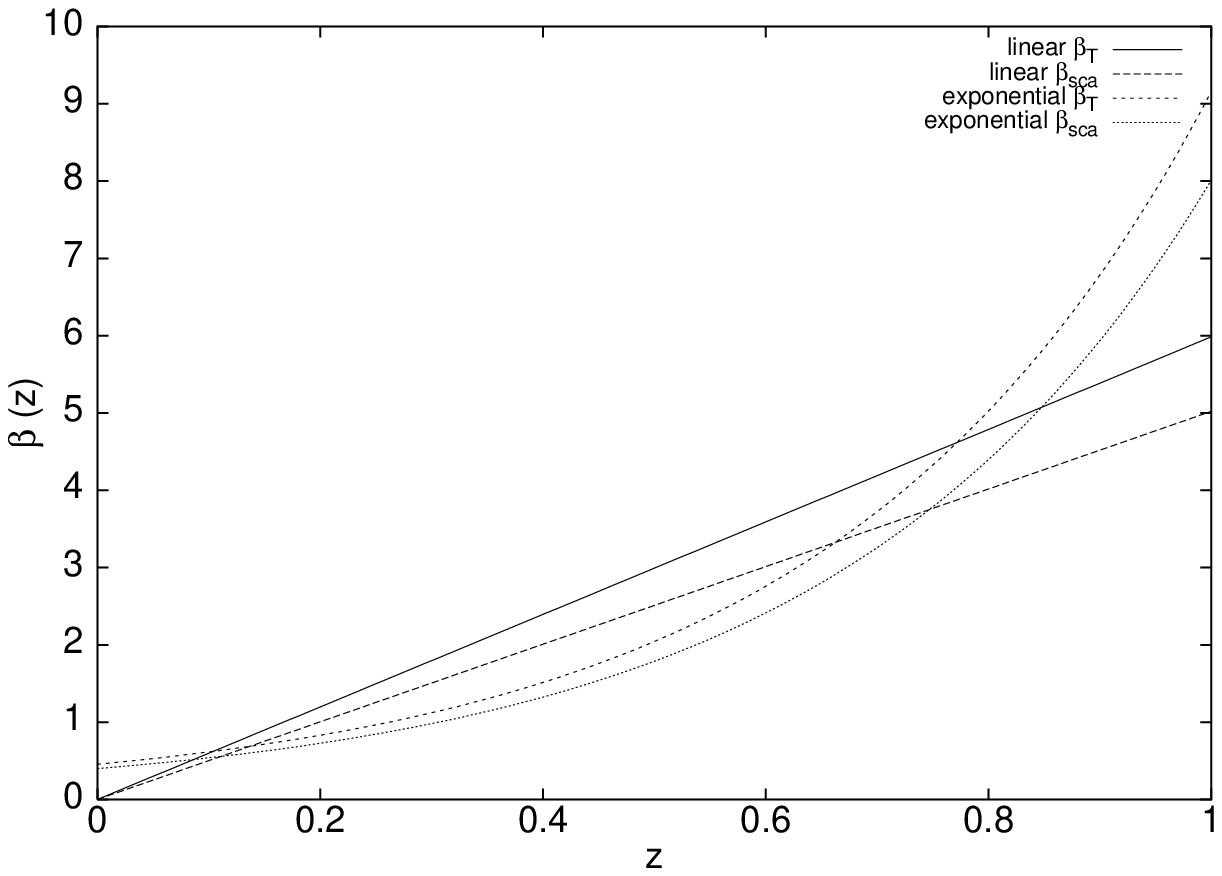} &
\includegraphics[width=0.45 \textwidth]{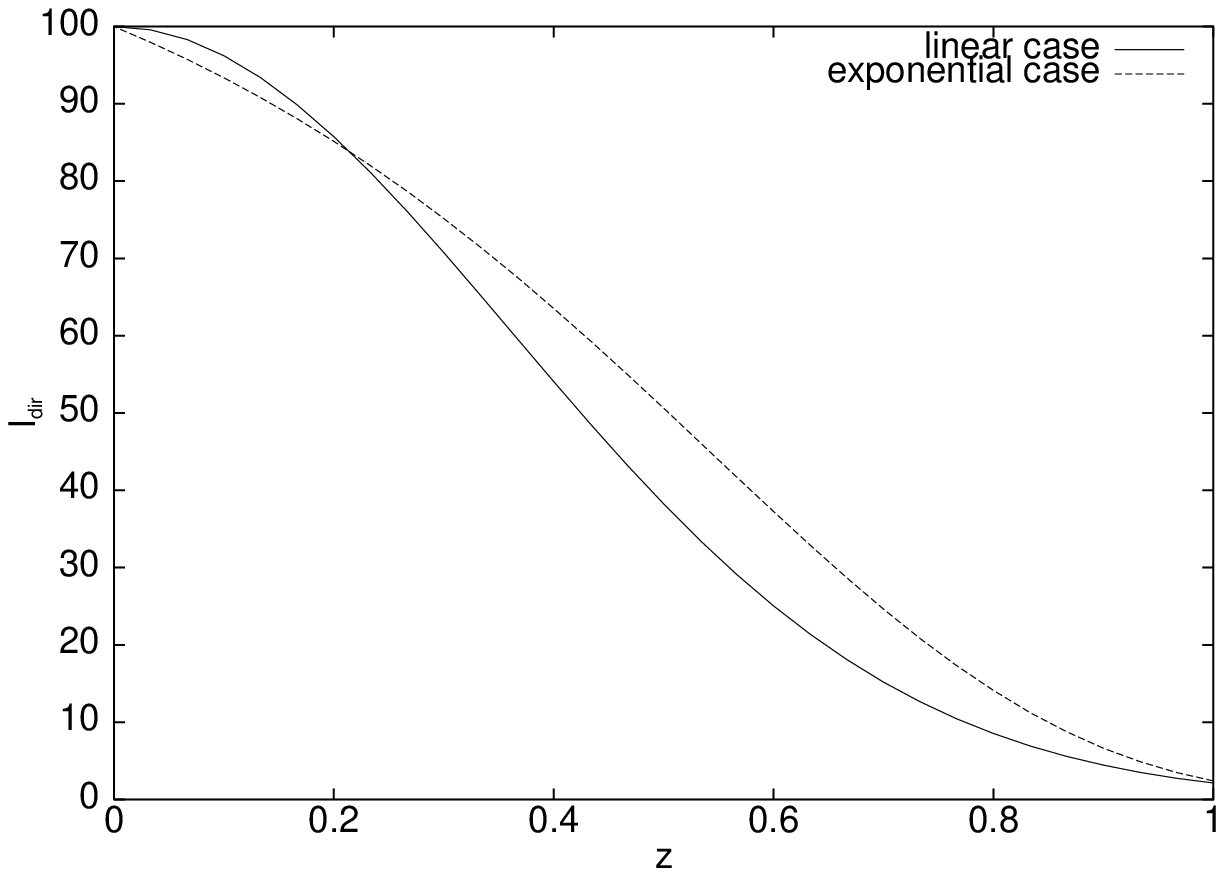} 
\end{tabular}
\caption{\label{beta}For the parameters given in the
  text, when $\tau\approx 3$: (I) absorption and scattering properties of the media in the
  case of linear and exponential height dependence, (II) direct intensity for the
  linear and exponential case as a
  function of height.}
\begin{tabular}{cc}

    (I) & (II)  \\
\includegraphics[width=0.45 \textwidth]{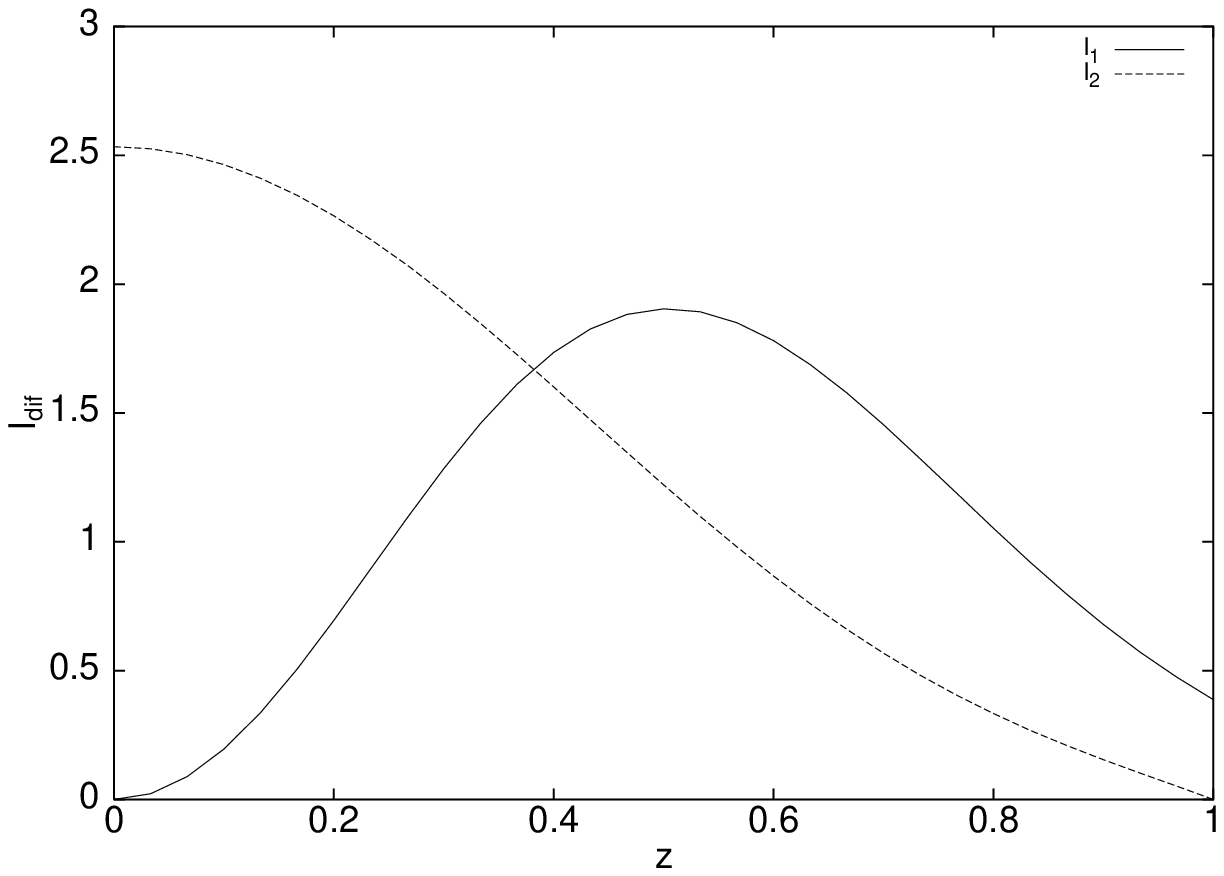} &
\includegraphics[width=0.45 \textwidth]{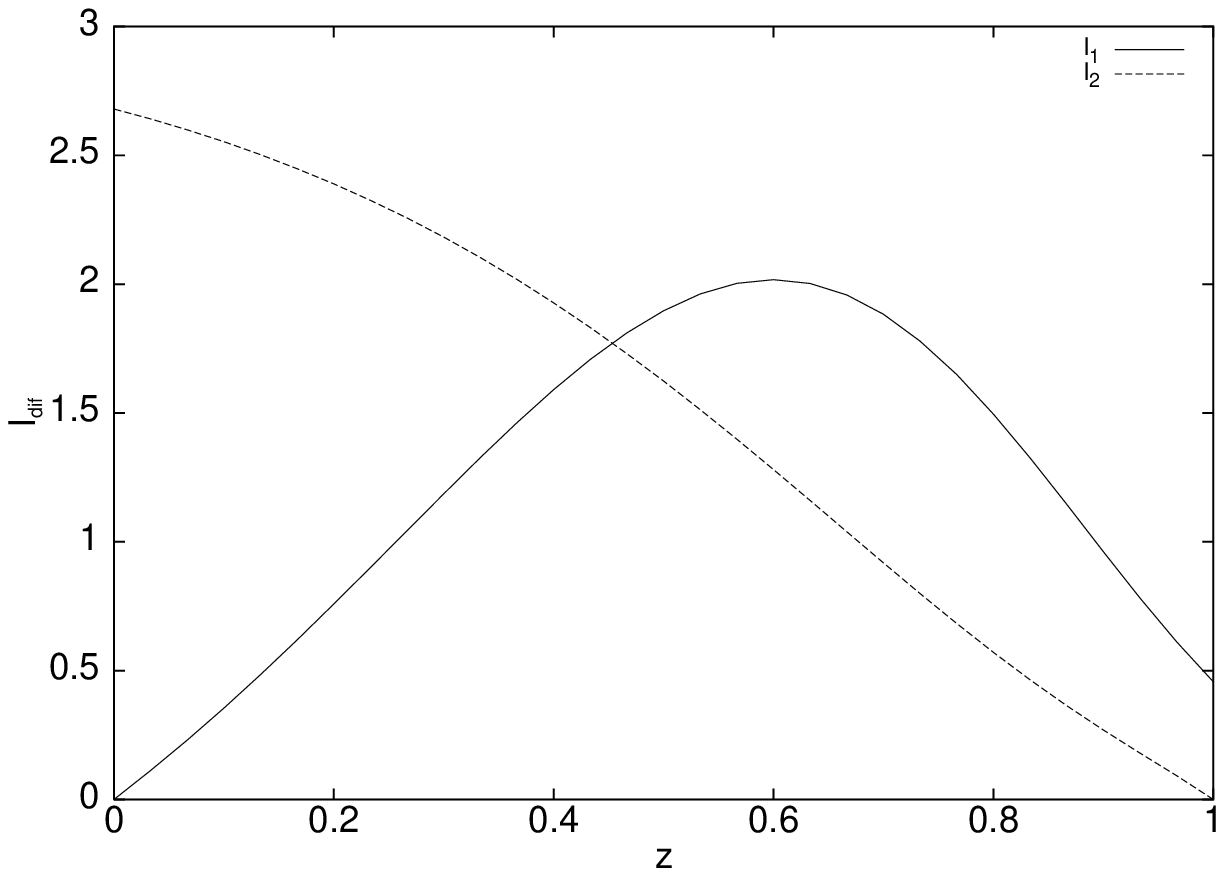} 
\end{tabular}
\caption{\label{exact}Exact two-stream components $I_1(z)$ (down) and $I_2(z)$
  (up) of the diffused intensity from the top $z=0$ to the bottom $z=1$ of the
  medium when the scattering and absorbing functions are linear (I) and exponential (II). In both cases the total optical path is $\tau_N\approx3$ and the incident direct radiation $I_0=100$.}
\end{figure}

The general solution for the downwards intensity is,

\begin{equation}
{I_1}(z) = e^{(e^{(a\,z)}\,b)}\,
\mathit{C_2} + e^{( - e^{(a\,z)}\,b
)}\,\mathit{C_1} - {\displaystyle \frac {( - c^{2} + a^{2}\,b^{2})\,(\mathit{
\mu_0}^{2}-\mu_0 \mu_2)}{16\,\pi \,( - c^{2}\,\mathit{\mu_2}^{2} + a^{2}\,b^{2
}\,\mathit{\mu_0}^{2})}}  \,I_0\,e
^{ \left( - \! \frac {c\,\mathit{\mu_2}\,e^{(a\,z)}}{\mathit{\mu_0
}\,a} \!  \right) } 
\end{equation}
and for the upwards intensity is,

\begin{equation}
{I_2}(z) = {\displaystyle \frac {c-ab}{c + a\,b}}\,e^{(e^{(a\,z)}\,b)}\,
\mathit{C_2} -  {\displaystyle \frac {c +a b}{ - c + a\,b}}  \,e^{( - e^{(a\,z)}\,b
)}\,\mathit{C_1} -  {\displaystyle \frac {( a^{2}\,b^{2}- c^{2} )\,(\mathit{
\mu_0}^{2}+\mu_0 \mu_2)}{16\,\pi \,(  a^{2}\,b^{2
}\,\mathit{\mu_0}^{2}- c^{2}\,\mathit{\mu_2}^{2})}}\,I_0\,e
^{ \left( - \! \frac {c\,\mathit{\mu_2}\,e^{(a\,z)}}{\mathit{\mu_0
}\,a} \!  \right) } 
\end{equation}
where $C_1$ and $C_2$ are arbitrary constants fixed with the boundary
conditions. The direct component of the intensity is,

\begin{equation}
I^{dir}(z)=I_0\,e^{\left(-\frac {c\,e^{a z} \mu_2}{a\,\mathit{\mu_0}}\right)}.
\end{equation}
For the specific choice $c=-0.8, a=3, b=0.2$, the constants $C_1$ and $C_2$ that satisfy the boundary conditions are $C_1 =-71.6787
$ and $C_2 =-8.51812e-05$. The total optical path, for this particular set
of parameters, is $\tau \approx 3$. The optical functions
$\beta_T(z)$ and $\beta_T(z)$ are shown in Fig. \ref{beta}-I and the direct
intensity in Fig. \ref{beta}-II. The two-stream components of the
diffused intensity are shown in Fig. \ref{exact}-II.

\section{The numerical method}\label{comparison}

In contrast to the examples discussed above, where the optical properties of
the media are known functions, in the general case the medium optical
properties are only known, based on measurements or models, at a discrete set of
points. In order to solve the radiative transfer equation, it is generally
assumed that these optical properties are piecewise constant in a layer around
the point, i.e. that the plane parallel layers are homogeneous
piecewise. As we shall see later this should not be assumed in the case of atmospheric layers, where
the optical properties show a strong (exponential) altitude dependence. 
We therefore propose to solve the problem numerically, using an adaptative step procedure
 and interpolating the optical values where necessary. 

Equation (\ref{radtrdir}) is an initial value problem and it is easily
integrated knowing the incident intensity on the top $I_0$. Equations
(\ref{diff1})-(\ref{diff2}) and their boundary conditions constitute a
two-point boundary problem which can be solved for instance with a shooting
method as those explained in \cite{nr,henar}. For problems with linear
boundary conditions, shooting methods are able to get the exact solution in
few steps. We choose values for all the
 variables at one boundary, which must be consistent with any boundary
 condition there, but are otherwise arranged to depend on
 arbitrary free parameters whose initial values are guessed. We then integrate the ordinary differential
 equations with initial value methods, arriving at the other boundary, and
 adjust the free parameters at the starting point that zeros the discrepancies
 at the other boundary.  Thanks to this procedure the problem is reduced to the
 solution of an initial value problem where no assumption of piecewise
 homogeneity is required. 

The numerical integration of the equation for the direct component (\ref{radtrdir}) and of the set of $n$ ordinary
 differential equations (\ref{eq0}) for the diffused streams is done with a fifth-order
  Runge-Kutta (RG) scheme with variable step \cite{RG}. Given a equation $\frac{dI_i}{dz}=f(z,I_i)$ where $I_i(z)$ is known, the
 RG scheme uses a
 weighted average of approximated values of $f(z,I_i)$ at several points within the
 interval $(z,z+dz)$ to obtain $I_i(z+dz)$. In contrast to other methods where
 piecewise homogeneity is required, this method takes into account the
 variation along
 the layer of integration of the diffused intensity streams, the direct component and  the medium characteristics.  
For the adaptative step control we use a
 "step doubling" technique: the local  error is estimated by comparing a solution obtained with a fifth-order scheme and the one obtained with a fourth-order method. The integrating
 step is halved if this error is above a desired tolerance. Next, given the tabulated function $\beta_{T,i}=\beta_T(z_i)$ and
$\beta_{sca,i}=\beta_{sca}(z_i)$ with $i=1...N$, we interpolate to obtain the new values at the locations required
by the  RG and the adaptative step doubling integration scheme. We have chosen a cubic
interpolating spline which allows the interpolating formula to be smooth in the first
derivative, and continuous in the second derivative, both within the interval
and at its boundaries \cite{nr}. 

As an example, next we solve
for the direct and diffused intensity in the two benchmark problems
explained in section \ref{examples}. For
problems with optical depths of the order of $\tau\approx 1$, standard
calculations based on the homogeneous layer approximation divide the medium
into $N=10$ homogeneous layers \cite{Liou}. We will analyze problems of
$\tau\approx 3$ and keep a similar ratio for our
comparisons, we will then increase the number of layers. The
tabulated optical properties
$\beta_T(z_i),\beta_{sca}(z_i)$ at $N$ equidistant $z_i$ points are given as
input.

\begin{figure}
\begin{tabular}{cc}

    (I) & (II)  \\

\includegraphics[width=0.45 \textwidth]{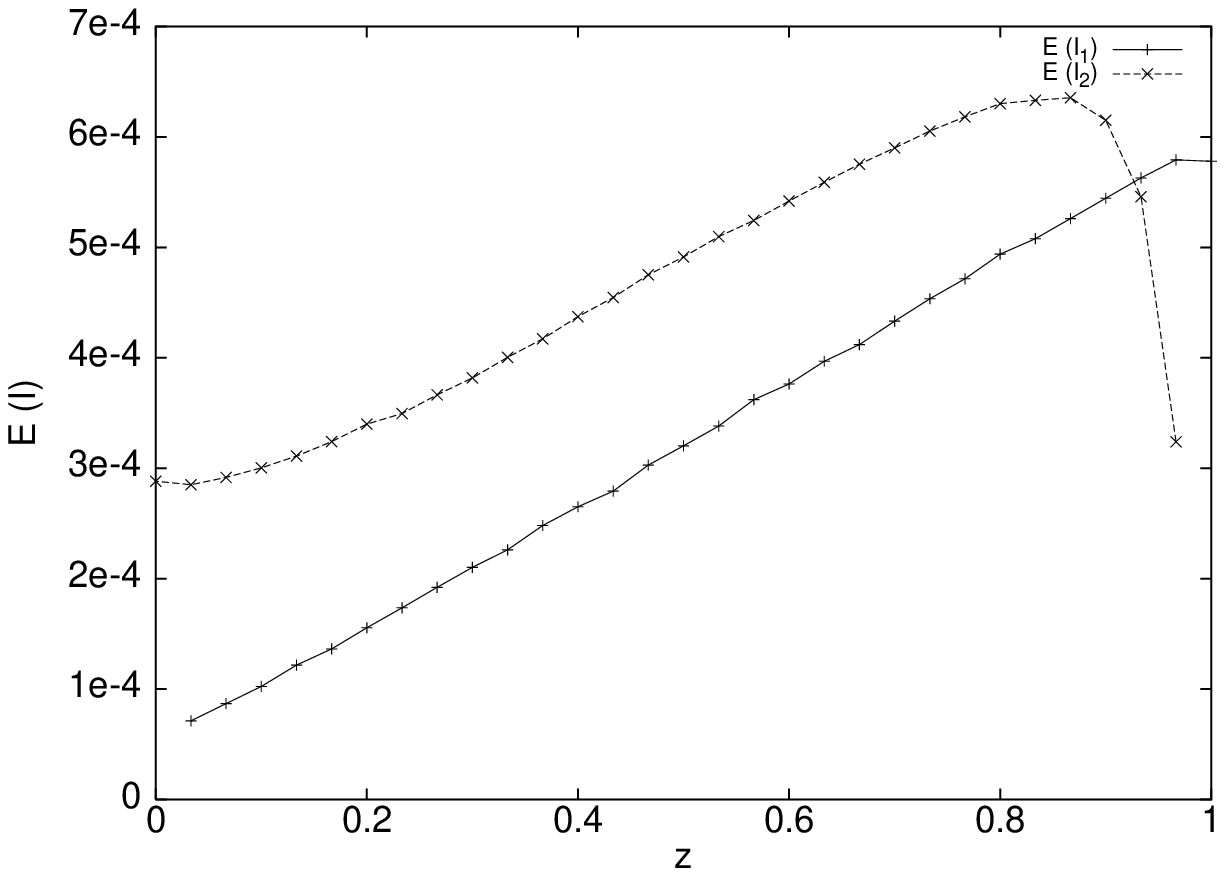} &
\includegraphics[width=0.45 \textwidth]{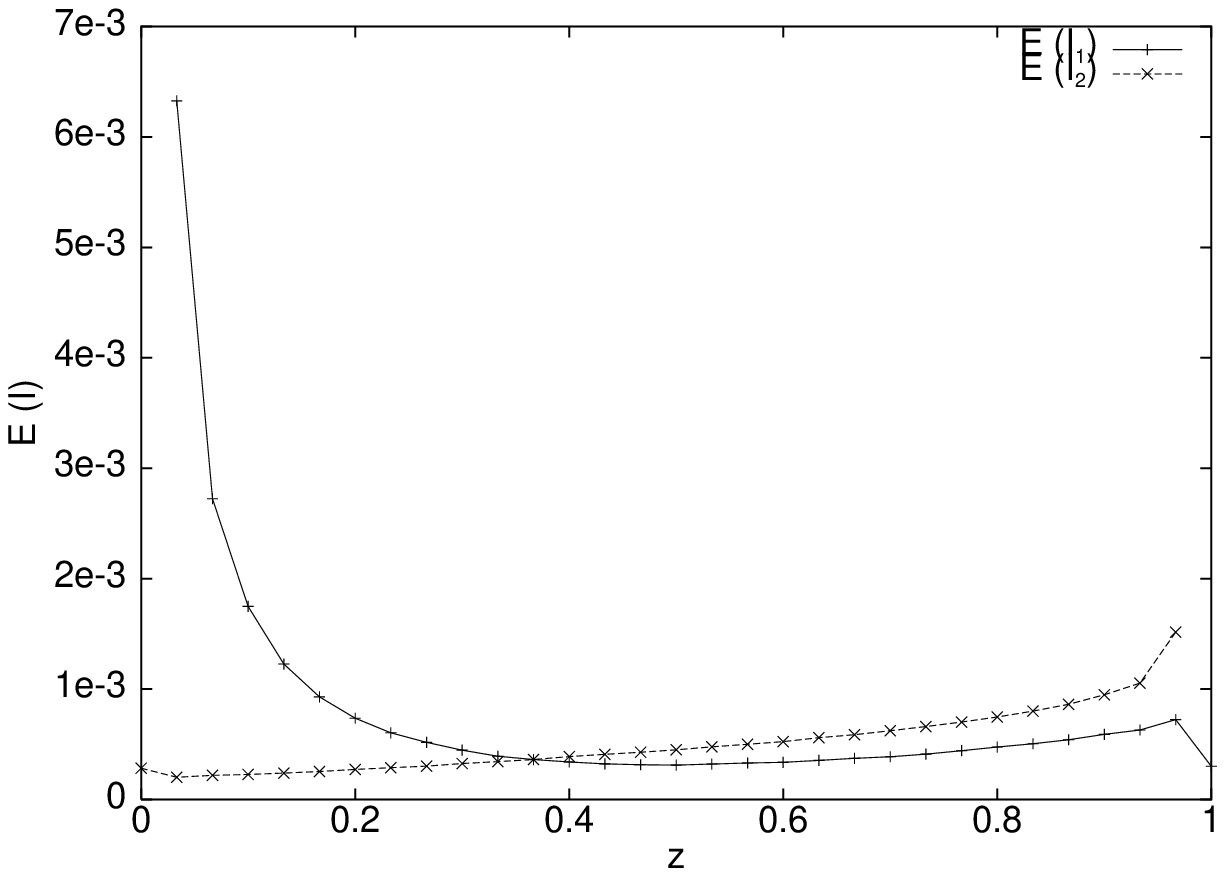}

\end{tabular}

\caption{\label{rel_error} Comparison of the numerical results with the exact
  solution of the diffused intensity for the
  case of a linear (I) and  exponential (II) optical functions. Relative local error
  $E (I_{1,2})=(I_{1,2}^{num}-I_{1,2}^{exact})/I_{1,2}^{exact}$.}
\end{figure}

A summary of the maximal
relative error $E(I)=|I_{dif}^{num}-I_{dif}^{exact}|/I_{dif}^{exact}$ for the benchmark problems as a function of the method and
number of initial layers is given in table \ref{table1}. We compare the relative
error in the solution
where piecewise homogeneity is assumed with the one obtained with our RG, step
doubling, interpolating method.  Figure
\ref{rel_error} shows the relative error as a function of height $z$ 
 for the step doubling
interpolating technique with $N=30$ layers, for the linear (I) and exponential
case (II). For a ratio of layers $N/\tau\approx 10$ the piecewise homogeneous approximation leads to errors of the
order of $10\%$ which can be reduced to $1\%$ by increasing this ratio to
$N/\tau\approx 80$. The step doubling
interpolating RG technique permits the adaption of the
numerical integration to the local characteristic length-scales of the medium
 and, for the same input, increases the accuracy of the
solution between one and two orders of magnitude allowing for fast and
accurate radiative transfer calculations. 

\begin{table}[htb]
\begin{center}
\caption{\label{table1}Solution of the linear and exponential problems with
  $\tau\approx 3$. Maximal local relative error in the diffused intensity as a function of the number of layers.}\vskip 1.5mm
\begin{tabular}{|c|c|c|c|}    \hline\hline
linear case &  N=30 & N=60    & N=240    \\ \hline
 piecewise homogeneous layers&  11$\%$ & 5.5$\%$  & 1.4$\%$  \\ \hline
 adaptative step interpolating method&  0.07$\%$ & 0.07$\%$  & 0.07$\%$    \\ \hline\hline
exponential  case & N=30 & N=60  & N=240    \\ \hline
 piecewise homogeneous layers&  15$\%$ & 8$\%$ & 2$\%$ \\ \hline
  adaptative step interpolating method &  0.63$\%$ & 0.16$\%$  & 0.09$\%$     \\
 \hline\hline
\end{tabular}
\end{center}
\end{table}

\section{Conclusions and discussion}\label{conclusion}

We have considered the discrete-ordinate approach to the radiation transfer
equation. In view of our calculations the error in the plane parallel
approach, assuming $N$ piecewise homogeneous layers, with
$N\approx10 \tau$ can go up to $10\%$ of the diffused intensity and thus the
inhomogeneities in the atmosphere cannot be ignored. We have validated a general purpose numerical method that,
based on an  adaptative step integration and an interpolation of the local
optical  properties,  can significantly improve (up to two orders of
magnitude) the  accuracy of the solution. This is furthermore of interest for
practical applications, such as atmospheric radiation transfer, where the scattering and absorbing properties
of the media are only known, based on experiments or theoretical models, at
certain discrete points.  Furthermore, this numerical method can be straightforward
extended to multiple streams, non trivial boundary conditions and non
uniform scattering function allowing for fast and
accurate solutions of the radiative transfer equation.

\section{Acknowledgments.}
M.-P. Z. is supported by the Instituto Nacional de Técnica Aerospacial (Spain).
A.M.M. thanks to the Spanish Government for a Ram\'on y Cajal Research Fellowship.


\begin{thebibliography}{9}

\bibitem{Stamnes}K. Stamnes. {\sl The Theory of Multiple Scattering of
    Radiation in Plane Parallel Atmospheres.} Reviews of Geophysics, Vol. 24,
  no. 2, pp. 299-310 (1986).

\bibitem{Liou} K.-N. Liou, {\sl Applications of the Discrete-Ordinate Method
    for Radiative Transfer to Inhomogeneous Aerosols Atmospheres.} Journal of
    Geophysical Research, vol. 80, no.24, pp. 3434-3440 (1975).

\bibitem{Kylling} A. Kylling, K. Stamnes, {\sl A Reliable and Efficient
    Two-Stream Algorithm for Spherical Radiative Transfer; Documentation of
    Accuracy in Realistic Layered Media.} Journal of Atmospheric Chemistry,
    vol. 21, pp. 115-150 (1995).

\bibitem{Carmen&Ana}C. C\'ordoba-Jabonero, L. M. Lara, A. M. Mancho,
  A. M\'arquez, R. Rodrigo {\sl Solar Ultraviolet transfer in the Martian
    atmosphere: biological and geological implications}. Planetary and Space
  Science, vol. 51, pp. 399-410 (2003).



\bibitem{nr} W.H.Press, S.A.Teukolsky, W.T. Vetterling, B. P. Flannery. {\sl
    Numerical Recipes in C}. Cambridge University Press 1994.


\bibitem{henar} V.M. P\'erez-Garc\'{\i}a, H. Herrero and J.J Garc\'{\i}a-Ripoll. {\it M\'etodos num\'ericos}, preprint.  http://matematicas.uclm.es/ind-cr/metmat/edovp.pdf. pp 13.6-13.7.



\bibitem{RG} Cash, J.R., and Karp, A. H., ACM {\sl Transactions on Mathematical software, vol. 16, pp. 201-202}.




\end{thebibliography}
\end{document}